\documentclass[preprint,12pt]{elsarticle}

\usepackage{amssymb}

\usepackage{graphicx}	
\usepackage{amsmath}	
\usepackage{amssymb}	
\usepackage{color}
\usepackage{hyperref}
\usepackage{textalpha} 

\newcommand{\lya}{Ly{\ensuremath{\alpha}} }
\newcommand{\HI}{\ensuremath{\textrm{H} \, \textsc{i}}}
\newcommand{\HII}{\ensuremath{\textrm{H} \, \textsc{ii}}}
\newcommand{\HeI}{\ensuremath{\textrm{He} \, \textsc{i}}}
\newcommand{\HeII}{\ensuremath{\textrm{He} \, \textsc{ii}}}
\newcommand{\HeIII}{\ensuremath{\textrm{He} \, \textsc{iii}}}

\journal{Astronomy and Computing}

\begin{document}

\begin{frontmatter}



\title{TensorFlow Hydrodynamics Analysis for Ly-$\alpha$ Simulations}


\author[Princeton]{Jupiter Ding}

\affiliation[Princeton]{organization={Department of Astrophysical Sciences},
            addressline={Princeton University, Peyton Hall}, 
            city={Princeton},
            postcode={08544}, 
            state={NJ},
            country={USA}}
\affiliation[LBL]{organization={Lawrence Berkeley National Laboratory},
            addressline={1 Cyclotron Road}, 
            city={Berkeley},
            postcode={94720}, 
            state={CA},
            country={USA}}

\affiliation[IPMU]{organization={Kavli IPMU (WPI), UTIAS, The University of Tokyo},
            city={Kashiwanoha},
            postcode={277-8583}, 
            state={Chiba},
            country={Japan}}
            
\author[LBL,IPMU]{Benjamin Horowitz}
\author[LBL]{Zarija Luki\'c}

\begin{abstract}
We introduce the Python program THALAS (TensorFlow Hydrodynamics Analysis for Lyman-Alpha Simulations), which maps baryon fields (baryon density, temperature, and velocity) to \lya optical depth fields in both real space and redshift space. Unlike previous \lya codes, THALAS is fully differentiable, enabling a wide variety of potential applications for general analysis of hydrodynamical simulations and cosmological inference. To demonstrate THALAS's capabilities, we apply it to the \lya forest inversion problem: given a \lya optical depth field, we reconstruct the corresponding real-space dark matter density field. Such applications are relevant to both cosmological and three-dimensional tomographic analyses of Lyman Alpha forest data.
\end{abstract}




\end{frontmatter}


\section{Introduction}
A significant portion of the universe's baryonic matter is made up of diffuse hydrogen gas located between galaxies, called the intergalactic medium (IGM). As light from distant sources passes through IGM, the neutral hydrogen absorb predominantly at the (rest-frame) \lya frequency; due to the cosmological expansion and associated redshifting, we observe this \lya absorption at a longer wavelength, depending on the distance of absorbing hydrogen. This collection of \lya lines creates the \lya ``forest'' in the quasar's spectrum. The forest is therefore a tracer of residual neutral hydrogen in the IGM, which itself traces the universe's matter density, and ultimately, its large scale structure \citep{McQuinn2016, Rauch2000}.

The structure of the neutral hydrogen in the universe traced by the \lya forest can be modelled via hydrodynamical simulations \cite{1994ApJ...437L...9C}. Unlike collisionless n-body simulations used to model the distribution of mass under the gravitational collapse, hydrodynamical simulations also follow the evolution of the baryon temperature, density, and velocity. Modern hydrodynamical simulations have been able to recover a number of statistics observed from the \lya forest, including the HI column
density distribution, the pixel flux distribution function and the flux
power spectrum, at levels necessary to distinguish and constrain cosmological models \cite{1995ApJ...453L..57Z,1996ApJ...457L..51H,2006MNRAS.365..231V}. Hydrodynamical simulations of the \lya Forest are a key element for the interpretation and analysis of next-generation survey data, including the Dark Energy Spectroscopic Instrument 
\cite{2016arXiv161100036D} and the Prime Focus Spectrograph \cite{2022arXiv220614908G}.  Ensembles of simulations varying model parameters are needed for both constraining cosmological parameters and understanding the details of reionization or the interplay between the IGM and galaxy formation. Over the past few years there has been significant development in accelerating/enhancing simulations and inference pipelines based on them \cite{2019ApJ...872...13W,2020MNRAS.497.4742K,2022arXiv220912931C,2023MNRAS.518.3754C,2023MNRAS.tmp..418A}. 

In this paper, we describe THALAS (TensorFlow Hydrodynamics Analysis for Lyman Alpha Simulations), a new analysis pipeline that computes \lya optical depth from hydrodynamical simulation outputs. Like existing analysis codes (e.g. \cite{friesen_2016}), THALAS can operate as a key component of any hydrosimulation post-processing analysis that involves optimization or parameter inference; e.g., in calculating statistics like power spectra and correlation functions. However, THALAS is fully differentiable; therefore, for a given optimization process, THALAS can greatly increase computational efficiency and still maintain accuracy. For example, in a Monte Carlo analysis for cosmological parameter inference, one can make highly informative updates via Hamiltonian Monte Carlo techniques if the derivatives of the final observed quantity are available. In recent years there has been an increased interest in dynamical forward modeling approaches (e.g. \cite{Horowitz2019,2021ApJ...906..110H,2019A&A...630A.151P}) which attempt to iteratively solve for the underlying initial density conditions of the universe given the oberserved data. In order for these optimization methods to be computationally tractable, derivative based methods are needed.

Many differentiable models utilize packages that have capabilities for automatic differentiation. Here, we use the functionality of TensorFlow \citep{tensorflow2015-whitepaper}, which is an interface for efficient machine learning on heterogeneous distributed systems. In particular, we perform operations with TensorFlow functions wherever possible and use TensorFlow's \texttt{GradientTape} class to track gradients throughout the \lya optical depth computation. This enables THALAS to be compatible with Graphics Processing Units (GPUs) as well as Tensor Processing Units (TPUs). A similar Python package is JAX \citep{jax2018github}), which is designed to automatically differentiate native Python and NumPy functions in a GPU- and TPU-compatible way.

This paper is organized as follows: In Section \ref{sec:results}, we describe the main components of THALAS and demonstrate its accuracy. In Section \ref{sec:application}, we explain how THALAS can be applied to the problem of \lya forest inversion, where its differentiability allows rapid inference in an otherwise almost intractable problem. In Section \ref{sec:conclusion}, we discuss potential future work and extension of the method.

\section{Data, Methods, and Results}
\label{sec:results}
The THALAS code \footnote{\url{https://github.com/Jupiter1994/lya-tf}} performs a mapping from baryon fields (the output of a hydrosimulation) to the \lya optical depth fields. Both TensorFlow's vectorization capabilities and compatibility with GPUs allows this computation to be done efficiently.
The baryon fields used are (relative) baryon density, temperature, and velocity parallel to the line-of-sight; notated as $\rho_b/ \bar{\rho}_b$ (or simply $\rho_b$), $T$, and $v_z$ (or $v_\text{para}$), respectively. The redshift-space optical depth ($\tau_\text{red}$) field depends explicitly on the $T$, $v_z$, and $n_{\HI}$ fields, while the real-space optical depth ($\tau_\text{real}$) field only depends explicitly on $T$ and $n_{\HI}$. 

\subsection{Data}

THALAS is generally tailored for cosmological hydrodynamical simulations that compute the aforementioned baryon fields; in this paper, we test THALAS on simulations generated by the Nyx code \citep{Almgren_2013}. Nyx is an adaptive mesh, cosmological hydrodynamics simulation code optimized to analyze the \lya forest signature in the diffuse intergalactic medium \citep{Lukic2014}. Each simulation treats the universe as a cubic box with periodic boundaries and produces ``snapshots'' of the universe at user-requested redshifts. Additionally, each simulation box can be split up into parallel, one-dimensional lines-of-sight (LOSs) or ``skewers.'' Based on the LOS orientation, one can derive the redshift-space (i.e. observable) optical depth fields. (The orientation is along the z-axis, which is why we use $v_z$ and $v_\text{para}$ interchangeably.)

The THALAS routine is adapted from (and benchmarked against) the C++ toolkit Gimlet, which generally analyzes the \lya forest and the IGM in cosmological simulations \citep{friesen_2016}. Various aspects of Gimlet's implementation make it difficult to make Gimlet's tools differentiable. This makes it unfeasible, if not computationally impossible, to use Gimlet as part of an optimization scheme that requires backpropagation.

\subsection{THALAS Workflow}

The first step in THALAS is to compute $n_{\HI}$, which can be determined from $\rho_b$ and $T$. In addition to solving for gravity and the Euler equations for fluid dynamics, Gimlet models six atomic species within the gas: \HI, \HII, \HeI, \HeII, \HeIII, and $\text{e}^-$. Assuming ionization equilibrium, the result is a system of equations that relates the number densities of the six species. Per \cite{Lukic2014}, the set of equations is:

\begin{equation}
  \begin{aligned}
    & \left( \Gamma_{e, \HI} n_e + \Gamma_{\gamma, \HI} \right) n_{\HI}
      = \alpha_{\rm r, \HII} n_e n_{\HII} \\[1.5mm]
    & \left( \Gamma_{e, \HeI} n_e + \Gamma_{\gamma, \HeI} \right) n_{\HeI}
      = \left( \alpha_{\rm r, \HeII} + \alpha_{\rm d, \HeII} \right)
      n_e n_{\HeII} \\[1.5mm]
    & \left[ \Gamma_{\gamma, \HeII} + \left(\Gamma_{e, \HeII}
      + \alpha_{\rm r, \HeII} + \alpha_{\rm d, \HeII} \right)
      n_e \right] n_{\HeII} \\[1.5mm]
    & \qquad = \alpha_{\rm r, \HeIII} n_e n_{\HeIII}
      + \left( \Gamma_{e, \HeI} n_e + \Gamma_{\gamma, \HeI} \right)
      n_{\HeI}
  \end{aligned}
  \label{eq:equil_species}
\end{equation}

In Equation~\ref{eq:equil_species}, the radiative recombination terms ($\alpha_r$), dielectronic recombination terms ($\alpha_d$), and collisional ionization rates ($\Gamma_e$), strongly depend on $T$. Furthermore, along with this system of equations, there are three closure equations for the conservation of charge and hydrogen and helium abundances.

We do not solve these equations directly in THALAS; rather, we take Gimlet's $n_{\HI}$ routine and interpolate it over the domain $\log_{10}\rho_b \in [-2, 3]$ and $\log_{10}T \in [3, 6]$ via SciPy's \texttt{scipy.interpolate.RectBivariateSpline} function. Afterwards, we apply the interpolated function to the Nyx $\rho_b$ and $T$ fields to get $n_{\HI}$. This allows rapid back-propogation in our analysis without having to differentiate through the solver of the non-linear system of equations. If variation of the UV background (i.e. $\Gamma_\gamma$) is of concern, this interpolation dimensionality can be expanded.

Once we have $n_{\HI}$, we can compute $\tau$ from $T$ and $n_{\HI}$. The formula for $\tau$ in real space is reproduced from \cite{Lukic2014} below:

\begin{equation}
\tau_\nu 
\equiv \int n_{\HI} \sigma_\nu dr \label{eq:tau_definition}
\end{equation}

Here, $\nu$ is the frequency, $\sigma_\nu$ is the cross-section of the \lya interaction, and $dr$ is the proper path length element. Assuming a Doppler line profile--which takes the shape of a Gaussian--this formula becomes:

\begin{equation}
\tau_\nu = 
\frac{\pi e^2}{m_e c} f_{12} \int \frac{n_{\HI}}{\Delta \nu_D} \frac{\text{exp} \left[-\left(\frac{\nu - \nu_0}{\Delta \nu_D}\right)^2 \right]}{\sqrt{\pi}} dr,
\end{equation}\label{eq:tau_real}

where $e$ is the fundamental electric charge, $m_e$ is the electron mass, and $c$ is the speed of light. Meanwhile, $\nu_0$ is the line center frequency, $f_{12}$ is the upward oscillator strength of the \lya resonance transition of $\nu_0$, and $\Delta \nu_D = (b/c) \nu_0$ is the Doppler width with the Doppler parameter $b = \sqrt{2k_B T/m_H}$.

To write the formula for $\tau$ in redshift space, one rewrites the path length as $dr = a dx = dv / H$, where $r$ is the proper distance, $a$ is the scale factor, $x$ is the comoving distance, $H$ is the Hubble expansion rate at the redshift of the particular simulation snapshot, and $v$ is the velocity coordinate (representing $v_z$). The resulting formula is:

\begin{equation}
\tau_v 
= 
\frac{\pi e^2 f_{lu} \lambda_0}{m_e c H} \int n_{\HI} \frac{1}{\sqrt{pi} b} \text{exp} \left[- \left(\frac{v - v_0}{b} \right)^2 \right] dv.
\end{equation}\label{eq:tau_red}

Similar to $f_{12}$ in Equation~\ref{eq:tau_real}, $f_{lu}$ is the oscillator strength for a particular line. $\lambda_0$ is the rest-frame line center and $v_0$ is the Hubble flow velocity at this redshift. The discretized version of this formula is

\begin{equation}
\tau_j
= 
\frac{\pi e^2 f_{lu} \lambda_0}{m_e c H} \sum_i n_{H I, i} \left[\text{erf}(y_{i - 1/2}) - \text{erf}(y_{i + 1/2}) \right] 
\end{equation}\label{eq:tau_red_discretized}

where $j$ is a particular pixel index and $i$ indexes over the pixels in the skewer. $y = (v_{z, j} - v_{z, i} - v_0)/b$ is the line center shift from the pixel velocity, scaled by the broadening scale. 

\begin{figure*}
    \centering
    \includegraphics[width=\textwidth]{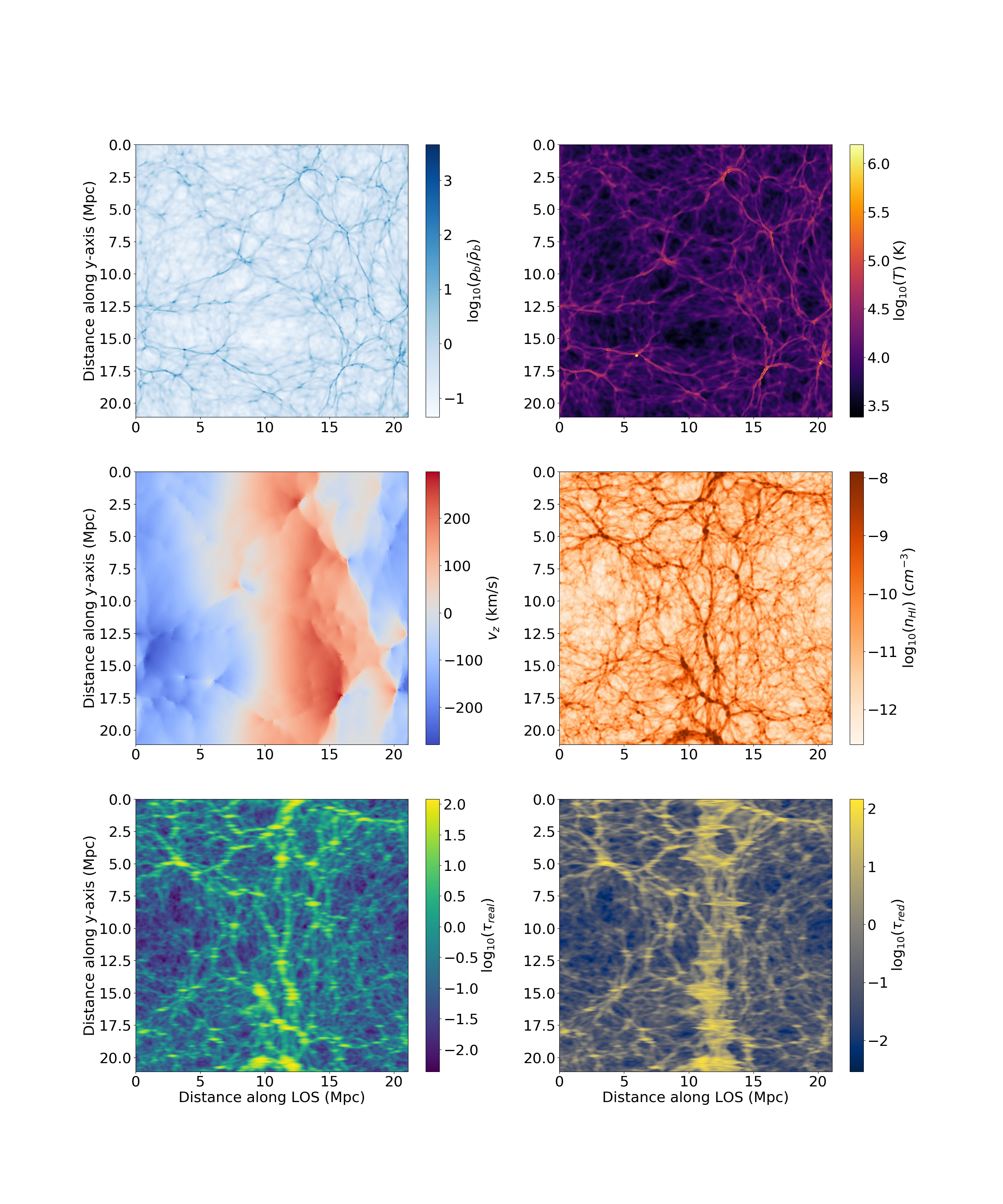}
    \caption{The fields at a 1024x1024 slice of a simulation snapshot at $z = 3$. The simulation-generated baryon fields are baryon density $\rho_b$, temperature $T$, and velocity $v_z$, while the fields computed by THALAS are number density of neutral hydrogen $n_{\HI}$, real-space optical depth $\tau_\text{real}$, and redshift-space optical depth $\tau_\text{red}$.}
    \label{fig:six_plots}
\end{figure*}

\subsection{Results}

\begin{figure}[t]
    \centering
    \includegraphics[width=0.8\columnwidth]{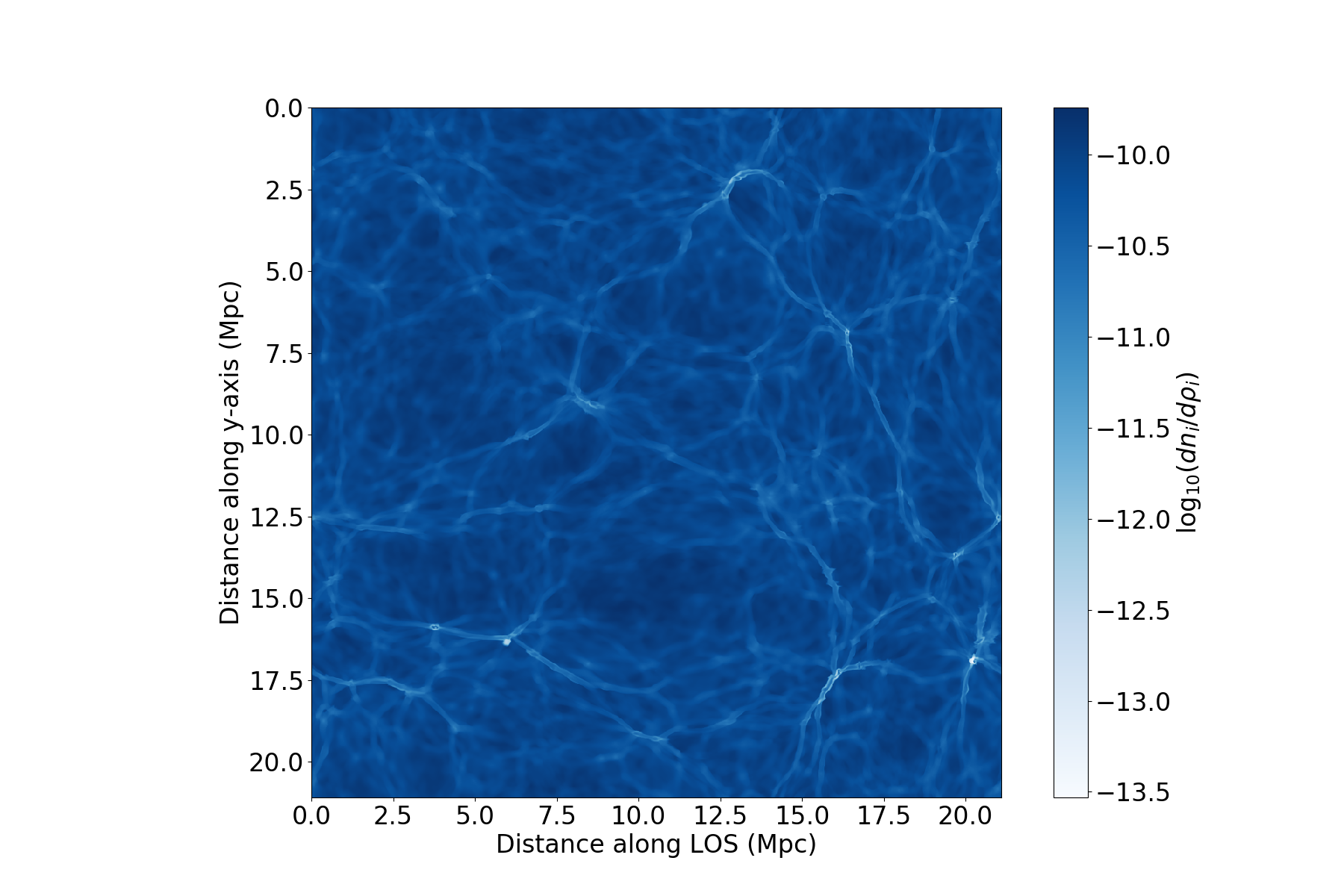}
    \caption{The $\frac{\partial n_{\HI}}{\partial \rho_b}$ field at the slice shown in Figure~\ref{fig:six_plots}. Neutral hydrogen is a biased tracer of the underlying dark matter field with nonlinear mapping. In dense regions, the response of neutral hydrogen to density becomes saturated, while in less dense regions it is more responsive.}
    \label{fig:dn_drho}
\end{figure}

\begin{figure*}
    \centering
    \includegraphics[width=\textwidth]{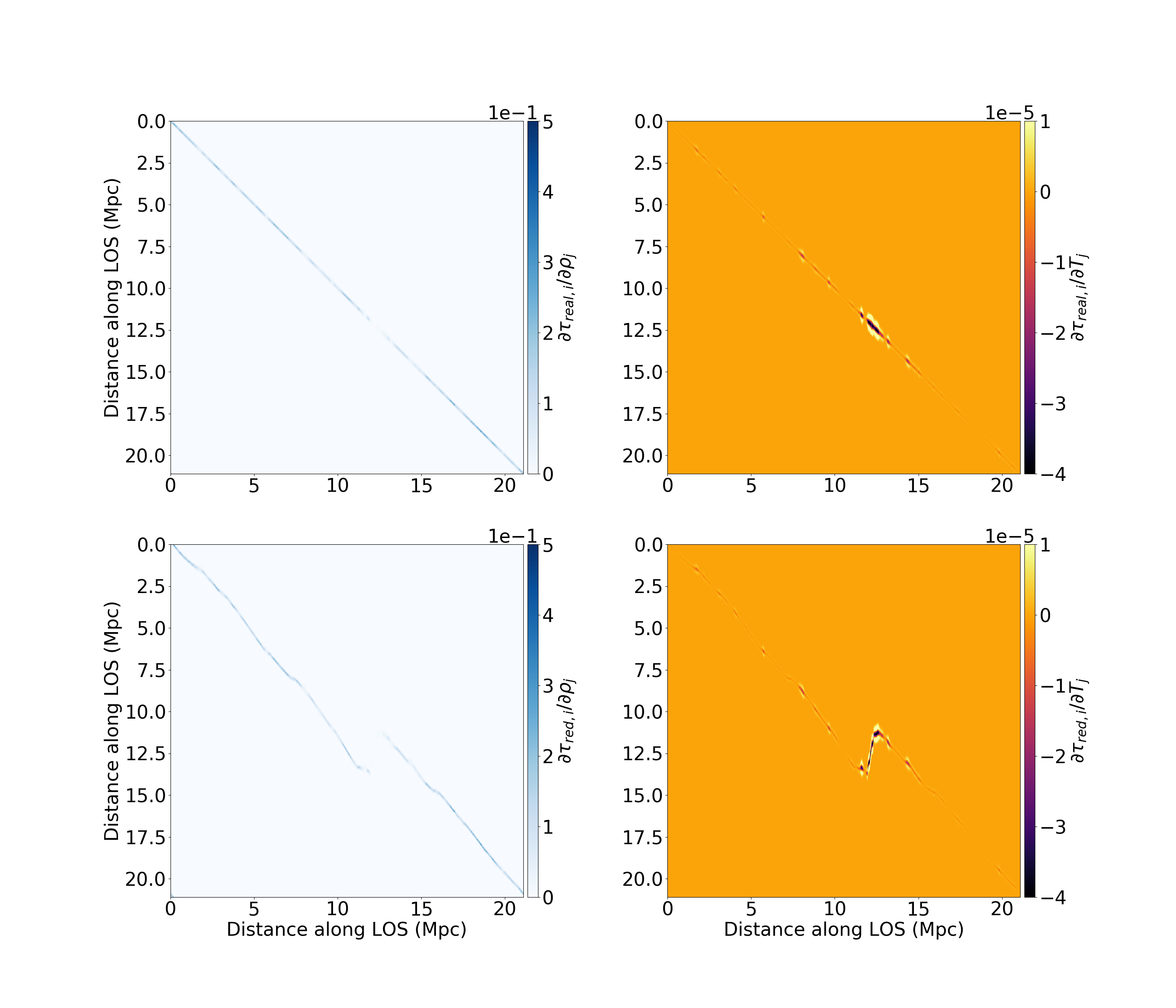}
    \caption{The Jacobians of $\tau_\text{real}$ and $\tau_\text{red}$ with respect to $\rho_b$ and $T$ along one skewer. Note that for $\tau_\text{real}$, the nonzero terms of the Jacobian are on or near the diagonal, but this is not the case for $\tau_\text{red}$. The redshift-space distortions are due to the nonzero $v_z$ field.}
    \label{fig:jacobians}
\end{figure*}

We test THALAS on the $z = 3$ snapshot of the simulation which contains $1024^3$ cells, and it has a length of 20 $h^{-1}$ Mpc along each edge. Taking a 1024x1024 slice from this snapshot, Figure~\ref{fig:six_plots} shows the baryon density, temperature and velocity along the line of sight, as well as the fields derived via THALAS: neutral hydrogen density and optical depth in the real- and redshift-space. We will further demonstrate that THALAS can extract the derivatives of the derived fields with respect to the baryon fields. Figure~\ref{fig:dn_drho} shows the partial derivative of $n_{\HI}$ with respect to $\rho_b$ throughout the aforementioned slice. Additionally, in Figure~\ref{fig:jacobians}, we take one skewer and calculate the Jacobian matrices of $\tau_\text{real}$ and $\tau_\text{red}$ with respect to $\rho_b$ and $T$. This shows the non-local nature of the redshift-space calculations vs. the real space.

\begin{figure}
    \includegraphics[width=\columnwidth]{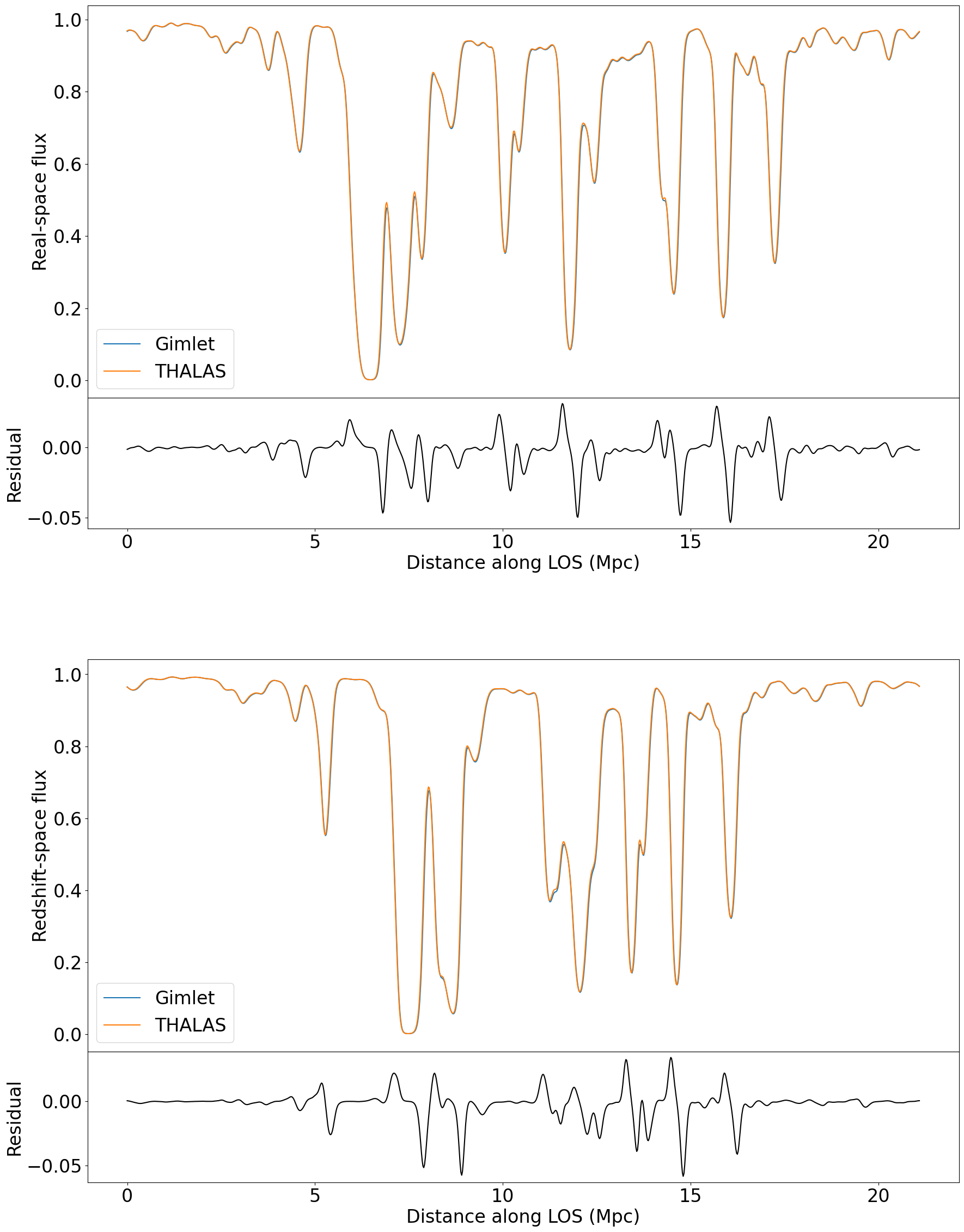}
    \caption{\textbf{Top:} Comparison of the real-space flux computed by Gimlet and THALAS along one line-of-sight. \textbf{Bottom:} Comparison of the redshift-space flux computed by Gimlet and THALAS along one line-of-sight.}
    \label{fig:comparison}
\end{figure}

\begin{figure}
    \includegraphics[width=\columnwidth]{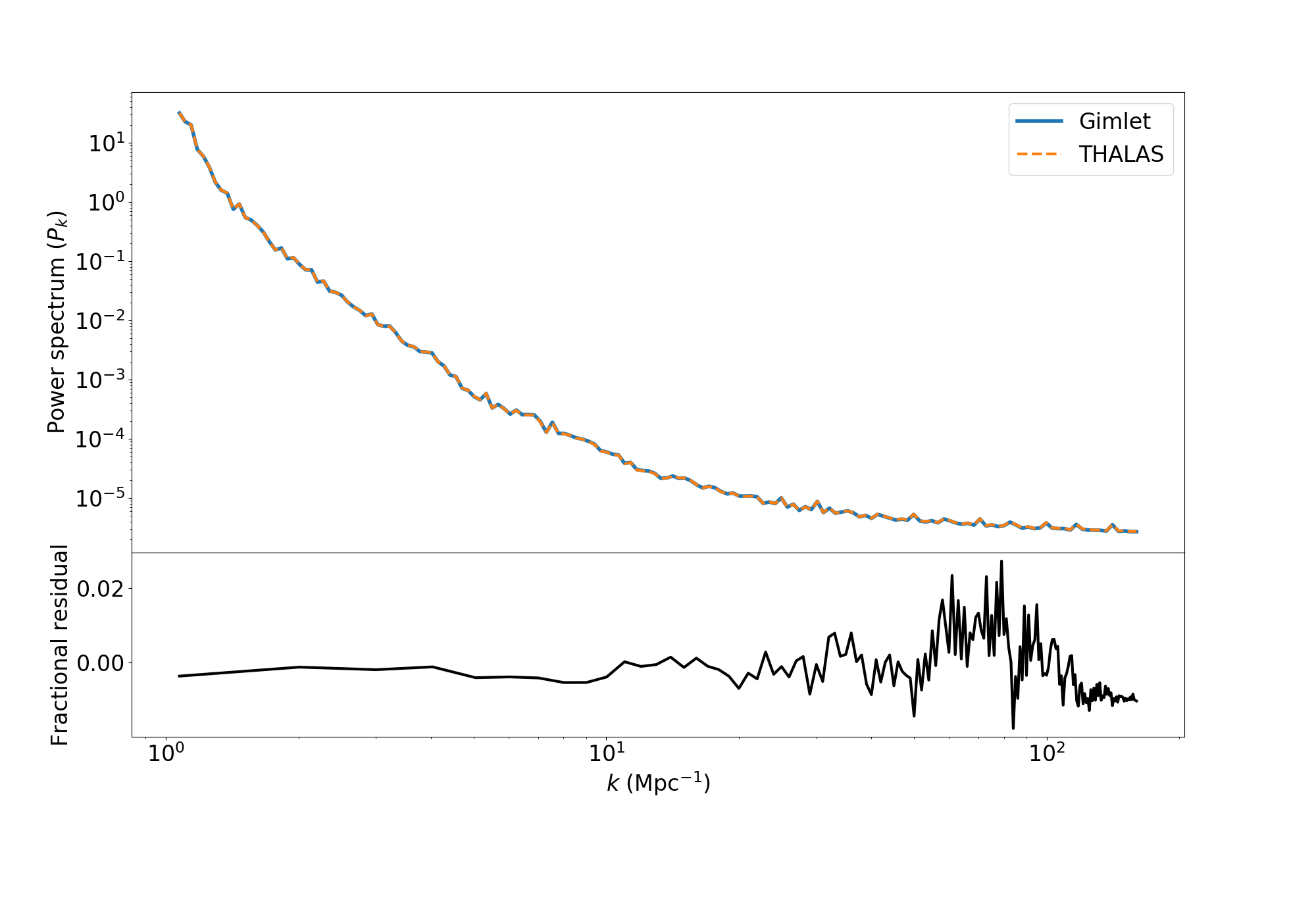}
    \caption{Comparison of the power spectrum analysis given the Gimlet- and THALAS-computed real-space optical depth fields for an arbitrary 1024x1024 slice.}
    \label{fig:comparison_ps}
\end{figure}

As a demonstration of THALAS's accuracy, Figure~\ref{fig:comparison} shows that Gimlet and THALAS's calculations for real- and redshift-space flux agree well along an arbitrarily-chosen skewer. We note that the computational complexity is essentially the same between Gimlet and THALAS; in practice, we translate Gimlet's C code into Python code with changes in implementation to allow memory efficient derivative calculations. We also perform a power spectrum analysis on the Gimlet- and THALAS-computed real-space optical depth fields for an arbitrary 1024x1024 slice in Figure~\ref{fig:comparison_ps}. Again, THALAS agrees extremely well with the Gimlet result, with the error being below $\sim$0.5\%. We believe this error is primarily due to both differences in integration schemes and numerical rounding errors.

\section{Application to Inversion of Lyman Alpha Forest}\label{sec:application}

To demonstrate the utility of THALAS's differentiable framework, we apply our technique to the \lya forest inversion problem: Given a \lya forest skewer, we want to reconstruct the real-space density field along the LOS associated with this skewer (e.g., \citealp{nusser_first_1999, 2001Pichon, croft_toward_2002, zaroubi_matter_2006}). This reconstruction can be utilized for cosmological analysis: If one evolves the dark matter density field backwards in time, they can subsequently reconstruct the initial dark matter density field, whose power spectrum contains all cosmological information. Additionally, density skewers can be used to tomographically reconstruct three-dimensional density maps to trace the underlying cosmic web, analogous to Wiener Filtered tomographic maps. 

In this section, we construct a simple differentiable model to map a dark matter density skewer (denoted $\rho$) to the associated baryon density and temperature. While this could be done with a neural network (e.g. \cite{2021harrington,horowitz2021hyphy}), we will use smoothed exponential mapping motivated by the results in \cite{2021harrington}, which found good agreement over a wide range of physical scales. Specifically, we rely on the Fluctuating Gunn-Peterson Approximation (FGPA), using the following formulas for $\rho_b$ and $T$, where $G$ is a Gaussian smoothing function \citep{hui_equation_1997, kooistra2022}: 
\begin{equation}
\rho_b = \rho_{b,0} \cdot G (\rho, \sigma_\rho)^{\gamma_\rho-1},\label{eq:rho_b}
\end{equation}
\begin{equation}
    T = T_0 \cdot G(\rho, \sigma_T)^{\gamma_T-1}.
\label{eq:temp}\end{equation}
We fit the parameters in Eqs.~\ref{eq:rho_b} and~\ref{eq:temp} a priori by training over the whole simulation volume. In particular, we use SciPy's \texttt{scipy.optimize.minimize} function, which uses a L-BFGS \citep{liu1989limited} quasi-Newtonian solver. We find the best-fit parameters to be $\rho_{b,0} \sim 1.2$, $\sigma_\rho \sim 3.3$, $\gamma_\rho \sim 1.9$, $T_0 \sim 17000$, $\sigma_T \sim 4.0$, and $\gamma_T \sim 1.4$. 

\begin{figure*}[t]
    \includegraphics[width=\textwidth]{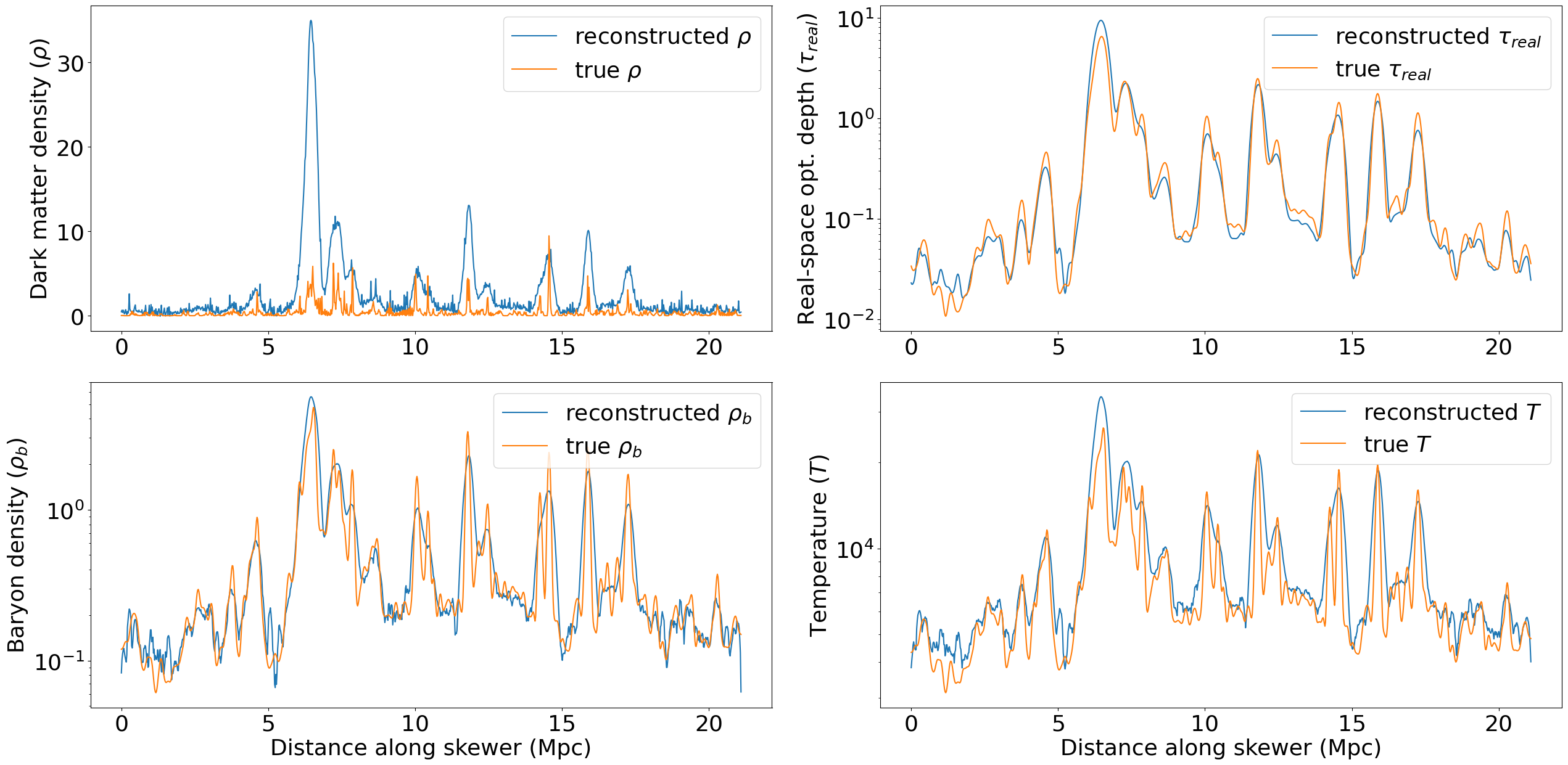}
    \caption{Applying THALAS to the \lya forest inversion problem: Given a real-space \lya optical depth ($\tau_\text{real}$) skewer, THALAS can reconstruct the dark matter density ($\rho$) skewer that underlies the $\tau_\text{real}$ field.}
    \label{fig:treal_recon}
\end{figure*}

Initially, we use random Gaussian noise as an initial guess for the dark matter skewer $\rho$. At each step in the optimization process, we use THALAS to compute the $\tau$ field associated with our guess for $\rho$, and we compare it to the given $\tau$ field. We use an MSE loss function, and we optimize this loss with a L-BFGS routine from the TensorFlow Probability (TFP) library. See Figure~\ref{fig:treal_recon} for results in reconstructing $\rho$ from a $\tau_\text{real}$ skewer. Even with a relatively limited forward model, we are able to recreate the general shape of $\rho$.

In order to reconstruct $\tau_\text{red}$, we would need to approximate $v_z$ given $\rho$. In theory, $v_z$ should be roughly proportional to the gradient of $\rho$ (e.g. \cite{zeldovich_gravitational_1970}). However, in the spatial scales studied in this work (i.e. for our 1024 by 1024 slice), the mapping from dark matter to velocity is nonlinear and contains significant sourcing terms outside of the skewer volume. While tomographic reconstructions could be used to gain insight to these modes \cite{Horowitz2019}, a full approximate inverse analysis that incorporates an estimation of $v_z$ is beyond the scope of this work. 

\section{Conclusions}
\label{sec:conclusion}
In this paper, we introduce THALAS, a TensorFlow-based model for mapping \lya forest observations to their underlying baryon fields. Because THALAS is fully differentiable, it has many potential applications for general analysis of hydrodynamical simulations. While we applied our tool to simulation outputs from Nyx, THALAS can also be applied to the outputs of other comparable simulations.

There are many potential applications of THALAS which we plan to explore in future work. Particularly, we plan to apply THALAS to the problem of dynamic forward-modelling, eventually combining THALAS with the HyPhy neural network model \citep{horowitz2021hyphy} to establish a new forward model. HyPhy is a painting method that synthesizes baryon fields from dark matter fields without performing a large-volume hydrodynamical simulation--therefore, it could replace our usage of FGPA in Section~\ref{sec:application}. Within this forward model, we would then use a third differentiable model to link the dark matter density fields to the initial density field, thus completing the full reconstruction process. 

Another potential application of THALAS would involve taking the \lya field and reconstructing the underlying power spectrum, rather than directly reconstructing the underlying matter density field. This has two sub-applications: one could either constrain cosmological parameters or FGPA parameters. In the former sub-application, one would set the cosmological simulation parameters and run the simulation. Afterwards, one could compare the simulated power spectrum with the power spectrum inferred by taking the simulated \lya field and reconstructing the power spectrum via FGPA and THALAS. In the latter sub-application, we would only run the cosmological simulation once, and we would vary the FGPA parameters. In both applications, differentability will allow efficient exploration of the posterior space with derivative-based methods.

\section{Acknowledgements}
This work was supported by the AI Accelerator program of the Schmidt Futures Foundation.



 \bibliographystyle{elsarticle-num} 
 \bibliography{cas-refs}





\end{document}